\documentclass[fleqn,twoside]{article}
\usepackage{espcrc2,cite,graphicx,amsmath,amsfonts,amssymb,mathrsfs}
\usepackage[figuresright]{rotating}

\newcommand{\ie}{{\it i.e.}}

\newcommand{\eg}{{\it e.g.}}

\newcommand{\cf}{{\it cf.}}
\newcommand{\etc}{{\it etc.}}

\newcommand{\fig}{Figure}

\newcommand{\Ref}{Ref.}
\newcommand{\Refs}{Refs.}
\newcommand{\Sec}{Section}

\newcommand{\Tab}{Table}

\newcommand{\stheta}{\sin^22\theta_{13}}
\newcommand{\deltacp}{\delta_{\mathrm{CP}}}
\newcommand{\ldm}{\Delta m_{31}^2}
\newcommand{\sdm}{\Delta m_{21}^2}

\newcommand{\figu}[1]{\fig~\ref{fig:#1}}

\newcommand{\bi}{\begin{itemize}}
\newcommand{\ei}{\end{itemize}}

\begin{document}

\title{
{\bf Neutrino Factory Superbeam}}

\author{{\large Patrick Huber},
\address[WIS]{{\it Department of Physics, University of Wisconsin,
       Madison, WI 53706, USA}}\thanks{E-mail: {\tt phuber@physics.wisc.edu}}
{\large Walter Winter},
\address[WUE]{{\it Institut f{\"u}r theoretische Physik und Astrophysik, 
Universit{\"a}t W{\"u}rzburg, D-97074 W{\"urzburg}, Germany}}\thanks{E-mail: {\tt winter@physik.uni-wuerzburg.de}}}

\begin{abstract}
\noindent {\bf Abstract}
\vspace{2.5mm}

We discuss the optimization of a neutrino factory for large $\sin^2 2 \theta_{13}$, where we 
assume minimum effort on the accelerator side. This implies that we 
use low muon energies for the price of an optimized detection system.
We demonstrate that such a neutrino factory performs excellent if combined with
the electron neutrino appearance channel. Instead of the platinum channel
operated with the muon neutrinos from the muon decays, we
propose to use the initial superbeam from the decaying pions and kaons, which might be
utilized at little extra effort. Since we assume out-of-phase bunches
arriving at the same detector, we do not require electron charge identification.
In addition, we can choose the proton energy such that we obtain a synergistic
spectrum peaking at lower energies.
We find that both the superbeam and the neutrino factory beam should used at the
identical baseline to reduce matter density uncertainties, possibly with the same
detector. This effectively makes the configuration a single experiment, which we 
call ``neutrino factory superbeam''. We demonstrate that this
experiment outperforms a low-energy neutrino factory or a wide band beam alone beyond
a simple addition of statistics.

\vspace*{0.2cm}
\end{abstract}

\maketitle

\section{Introduction}

Future neutrino oscillation facilities are primarily targeted towards the discovery of a non-zero $\stheta$,
the neutrino mass hierarchy determination, and the measurement of leptonic CP violation. All of these
measurements depend heavily on the magnitude of $\stheta$.
Two examples for experiments under way, which will test the magnitude of $\stheta$, are the T2K superbeam (SB) project~\cite{Itow:2001ee} and the Double Chooz reactor experiment~\cite{Ardellier:2006mn}.
From all such neutrino oscillation experiments, one may expect a $\stheta$ sensitivity limit of about $0.01$
in a decade from now (see, \eg, \Ref~\cite{Huber:2004ug}), \ie, $\stheta$ can either be constrained to
 be smaller than this value, or it will be discovered until then. Since the discovery of $\stheta$ affects the
experimental strategy how to measure the mass hierarchy and CP violation, this ``branching point''~\cite{Albrow:2005kw} will, at the latest, tell us how to proceed. 

If $\stheta$ turns out to be small,
one may optimize for a maximum $\stheta$ reach of all measurements. It seems that a neutrino factory~\cite{Geer:1997iz,Barger:1999jj,Cervera:2000kp,Huber:2002mx} (NF) or a higher gamma beta beam~\cite{Burguet-Castell:2003vv,Burguet-Castell:2005pa} could be the most promising alternative; for the optimization discussions, see, \eg,
\Refs~\cite{Huber:2005jk,Huber:2006wb}. For large $\stheta$, the measurement far away from the oscillation
maximum and matter density uncertainties affect the NF performance~\cite{Huber:2002mx,Ohlsson:2003ip}. 
Therefore, other alternatives may outperform a NF,
such as SB upgrades or beta beams (see, \eg, \Refs~\cite{Huber:2005jk,Barger:2007jq}
for a direct comparison).

Recently, potential improvements in the muon charge identification (CID) 
at low energies (see, \eg, \Ref~\cite{Ellistalk}) 
have brought back a NF into the discussion for large $\stheta$. In \Ref~\cite{Geer:2007kn},
a muon energy of $4.12 \, \mathrm{GeV}$ was proposed at baselines of $1 \, 000$ to $1 \, 500 \, \mathrm{km}$, an option similar to a wide band beam in terms of neutrino energies and baseline. The relatively low
muon energy allows for a significant reduction of the effort on the accelerator side, which means that
this option might indeed be competitive to SB upgrades, \etc. Other alternatives to improve the
NF for large $\stheta$ are the use of the magic baseline~\cite{Huber:2003ak,Huber:2006wb}, or
the platinum channel $\nu_\mu \rightarrow \nu_e$ (with the $\nu_\mu$'s coming from the muon decays), which requires electron detection with CID~\cite{Huber:2006wb}.
In this way, the impact of matter density uncertainties can be significantly reduced as well.

In this letter, we discuss the optimization of a low-energy NF. We compare such an
optimized experiment to wide band beams and beta beams. Furthermore, we propose to use the
neutrinos from the secondary pion/kaon beam in addition to the NF beam.
This SB, possibly targeted towards the same detector, will provide similar information
to the platinum channel. However, it has two major advantages: First, the (at least in 
a magnetized iron calorimeter) very difficult electron CID is not required if operated out-of-phase with the NF beam. Second, the proton energy can be chosen such that the 
event rates are peaking at lower energies, where the NF has less events.
We discuss the requirements for such a hybrid configuration, which we call ``neutrino factory superbeam''
({\sf NF-SB}), and we show the physics potential.

\section{Setup and Experimental Requirements}
\label{sec:setup}

\begin{figure*}[t!]
\begin{center}
\includegraphics[width=0.8\textwidth]{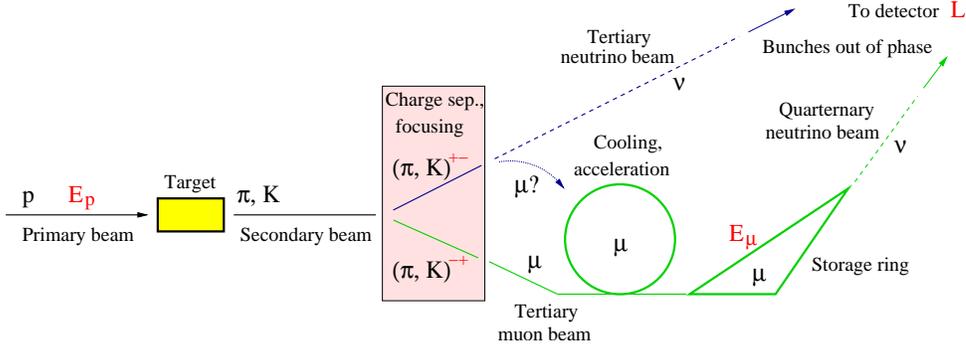}

\vspace*{-1.0cm}

\end{center}
\caption{\label{fig:schematics} Schematics of a {\sf NF-SB} (not to scale). Our
degrees of freedom are given by red/gray labels. Here we adopt the conservative
point of view that only half the muons can be collected for the NF.}
\end{figure*}

The primary information from a NF comes from the ``golden'' $\nu_e \rightarrow \nu_\mu$ appearance channel~\cite{Cervera:2000kp}.\footnote{Note we will also use antineutrinos in all of the discussed channels.} The two major challenges for large $\stheta$ are the matter density uncertainties and the lack of events at low energies, where the interesting higher-order oscillation effects are present. As demonstrated in \Ref~\cite{Huber:2006wb}, complementary information from the T-conjugated $\nu_\mu \rightarrow \nu_e$ appearance channel (platinum channel) would help, partly because of the correlated matter effect. However, CID for electrons in a magnetized iron calorimeter appears to be difficult 
because the electrons produce showers early (though it may not be impossible for low neutrino energies). In this letter, we propose
to use the SB from the pion and kaon decays instead of the platinum channel. The schematics for
such a setup can be found in \figu{schematics}. We require that it be operated out-of-phase with the NF beam, which means that CID will not be required for this channel. 
In addition, the proton energy can be adjusted 
within the range acceptable for the NF\footnote{The acceptable range is relatively wide $8 \, \mathrm{GeV} \lesssim E_{\mathrm{p}} \lesssim 30 \, \mathrm{GeV}$ (see, \eg, \Ref~\cite{Zismantalk}).} such that a spectrum peaking at lower energies can
be obtained. As one can see from \figu{schematics}, there are two major requirements for such a setup: First, the geometry has to be chosen such that the tertiary and quarternary neutrino beams point towards the same detector (not to scale). Second, the target station has to be able to separate and focus the pion and kaon charges in a way that both experiments can be operated simultaneously. From \figu{schematics}, this requirement may be acceptable if we only use half of the luminosity of an exclusive NF (because only one polarity can be operated at the same time). However, one may speculate that a more sophisticated target station might recycle a fraction of the muons from the SB mode. Note that, for the NF plus SB, we have now three phenomenological degrees of freedom: $E_p$, $E_\mu$, and $L$.

The main prerequisite for a low-energy NF is an improved detection system with sufficient
golden channel efficiencies at low energies. These efficiencies with a threshold possibly as low as $0.5 \, \mathrm{GeV}$
might be obtained by cuts leading to a higher charge mis-identification background, or a more refined
detector than a magnetized iron calorimeter (such as a NO$\nu$A-like detector or a liquid argon detector
with a large magnetic volume). For large $\stheta$,
such a background will not be very important as long as the background level is $\lesssim \stheta$. 
We follow \Refs~\cite{Huber:2006wb,Geer:2007kn} by using a  
low-threshold ($500 \, \mathrm{MeV}$) with improved energy resolution option ($15\%/\sqrt{E}$),
which we call {\sf Golden*}. We us 50\% efficiency for the golden channel requiring CID, and 90\% efficiency for the disappearance channel without CID. 
Note that an even higher appearance efficiency may be obtained for the price of larger backgrounds.
We use a total luminosity of $5 \, \mathrm{yr}$ running time  $\times$ $10^{21}$ useful muon decays/year $\times$ $50 \, \mathrm{kt}$ detector mass in each polarity ($\mu^+$, $\mu^-$ stored) if operated as NF alone, and half of that for the {\sf NF-SB}.  We assume that the number of useful muon decays/year approximately requires a target power of $4 \, \mathrm{MW}$, but this number scales with non-trivial aspects of the muon
cooling and acceleration system. For the comparison to the platinum channel, we again follow the optimistic choice in \Ref~\cite{Huber:2006wb} using 40\% efficiency, no upper threshold, and 1\% charge mis-identification (\cf, \eg,  \Ref~\cite{Rubbia:2001pk} for liquid argon).

For the SB operation, we have tested a {\sf MiniBOONE}-like wide band beam ($E_p=8 \, \mathrm{GeV}$) and an {\sf AGS}-like wide band  beam ($E_p = 28 \, \mathrm{GeV}$). Since the {\sf MiniBOONE}-like beam has turned out too have too low neutrino energies for our application, we focus on the {\sf AGS}-like beam in the following (see \Refs~\cite{Barger:2006vy,Barger:2006kp,Barger:2007jq}
for details on the simulation). Note that a considerably higher proton energy may lead to a significant loss
of efficiency for the NF. We assume an electron 
detection efficiency of 80\% without CID (in {\sf Golden*}) and that the experiments is subject to the same
NC background as the muons. This assumption is probably too optimistic, but the intrinsic beam background is larger anyway. Our systematical errors are chosen to be 5\% except for the signal normalization errors
in the NF ($2.5\%$). 
\begin{figure}[t!]
\begin{center}
\includegraphics[width=\columnwidth]{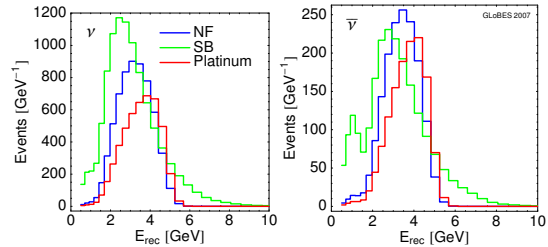}

\vspace*{-1.0cm}

\end{center}
\caption{\label{fig:rates} Appearance event rates as a function of the reconstructed neutrino energy.  
The figures compare the NF (golden), 
SB, and platinum channel appearance.
 The curves are computed for $E_\nu=5 \, \mathrm{GeV}$, $E_p=28 \, \mathrm{GeV}$, $L=1 \, 250 \, \mathrm{km}$, $\stheta=0.1$, $\deltacp=0$, $m_{\mathrm{Det}}=50 \, \mathrm{kt}$, $2.5 \, 10^{21}$ useful muon decays, and $P_{\mathrm{target}}=4 \, \mathrm{MW}$.}
\end{figure}
We compare in \figu{rates} the event rate spectra from the three potentially
used appearance channels (NF, SB, platinum).
The peak of the platinum channel is at higher energies than the peak of the
NF because of the corresponding spectra from the muon decays (the electron neutrinos
peak at lower energies than the muon neutrinos). On the other hand, $E_p$ is chosen
such that the SB peaks at lower energies, where oscillation effects are more
prominent (\cf, right panel). Note that this figure assumes the muon decay numbers of the {\sf NF-SB},
whereas an exclusive NF operation will allow for the double luminosity.

\begin{figure*}[t!]
\begin{center}
\includegraphics[width=0.75\textwidth]{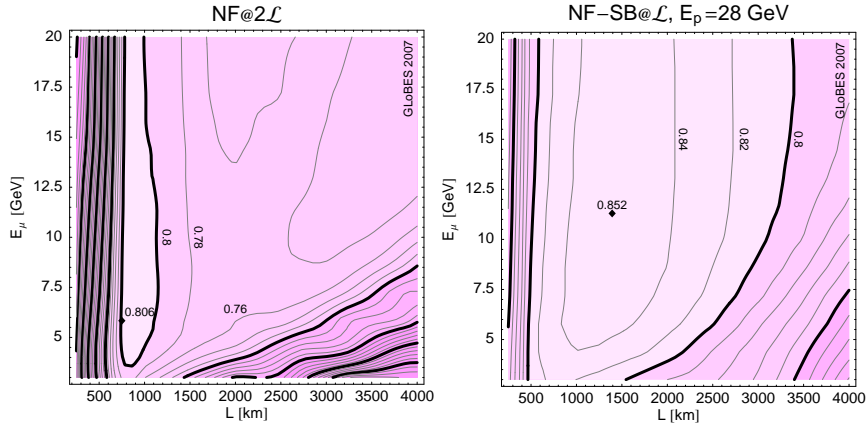}

\vspace*{-1.0cm}

\end{center}
\caption{\label{fig:optcomp} Fraction of $\deltacp$ for which CP violation can be measured
as a function of $L$ and $E_\mu$ ($\stheta=0.1$, $3 \sigma$ confidence level). The left panel is for a NF alone with twice the
luminosity as in the right panel. The right panel is for a NF plus SB combination
with a fixed proton energy of $E_p = 28 \, \mathrm{GeV}$. The contours are spaced by 0.02, where the
thick curves correspond to a fraction of $\deltacp \in \{0.8, 0.7, 0.6, \hdots \}$. The optima
are marked by the diamonds. The luminosity ``$\mathcal{L}$'' stands for $0.5 \, 10^{21}$ useful muon 
decays/yr.}
\end{figure*}

For the simulation, we use the {\sf GLoBES} software~\cite{Huber:2004ka,Huber:2007ji}. 
For the oscillation parameters, we use $\sin^2  \theta_{12}=0.3$, $\sin^2  \theta_{23}=0.5$, $\sdm = 7.9 \, \cdot 10^{-5} \, \mathrm{eV^2}$, $\ldm = 2.5 \, \cdot 10^{-3} \, \mathrm{eV^2}$, and a normal mass hierarchy unless stated otherwise (see, \eg, \Ref~\cite{Schwetz:2006dh}). In addition, we assume that $\sdm$, $\theta_{12}$, and the matter density will be known at the level of 5\%.

\section{Optimization for Large $\boldsymbol{\stheta}$}

As far as the optimizations of a low-energy NF and the {\sf NF-SB} are concerned, 
let us first of all focus on the NF alone. In \figu{optcomp}, left panel, we show the 
fraction of (true) $\deltacp$ for which CP violation can be measured
as a function of $L$ and $E_\mu$ (for $\stheta=0.1$). At the optimum (diamond), one can establish
CP violation for 81\% of all possible values of $\deltacp$. One has close-to-optimum performance
for $L \sim 800 \, \mathrm{km} - 1 \, 300 \, \mathrm{km}$ and $E_\mu \gtrsim 4 \, \mathrm{GeV}$,
which is consistent with the configuration discussed in \Ref~\cite{Geer:2007kn}.
For the mass hierarchy, we find that $L \gtrsim 800 \, \mathrm{km}$ to establish the mass hierarchy
for any $\deltacp$ (the $E_\mu$-dependence is moderate). In addition, note that for such a large $\stheta$, none of the discussed facilities has a problem to establish a non-zero $\stheta$. 

In \figu{optcomp}, right panel, we show the same optimization for {\sf NF-SB} with a fixed proton energy of $E_p = 28 \, \mathrm{GeV}$ and half the useful number of muon decays from the last example. 
At the optimum (diamond), one can establish CP violation for 85\% of all possible values of $\deltacp$.
The optimal baseline range depends on $E_\mu$. For instance, one has close-to-optimum performance for $L \sim 800 \, \mathrm{km} - 1 \, 500 \, \mathrm{km}$ for $E_\mu \gtrsim 4 \, \mathrm{GeV}$. For the mass hierarchy, shorter baselines $L \gtrsim 500 \, \mathrm{km}$ are, in principle, sufficient. This effect has been discussed in \Ref~\cite{Schwetz:2007py}. Since the NF luminosity in this hybrid is only half of the one for the NF alone, we conclude that we have a identified a synergy between the NF and the SB which goes beyond the simple addition of statistics (for a definition of synergy, see \Ref~\cite{Huber:2002rs}). Note that the SB alone operated at double luminosity
approximately compares to the low-energy NF alone (for $\stheta=0.1$ and the same detector).

\begin{figure}[t!]
\begin{center}
\includegraphics[width=0.7\columnwidth]{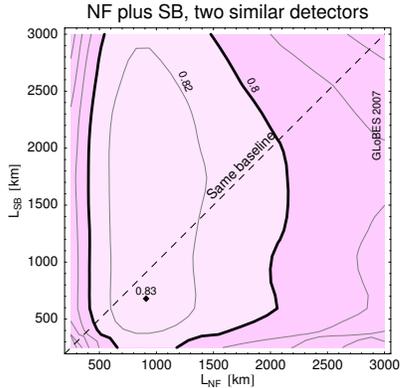}

\vspace*{-1cm}

\end{center}
\caption{\label{fig:twobase} Two-baseline optimization
of the NF plus SB with uncorrelated matter effect: Fraction of $\deltacp$ for which CP violation can be discovered as a function of $L_{\mathrm{NF}}$ and $L_{\mathrm{SB}}$ ($3 \sigma$). The muon energy is $E_\mu = 5 \, \mathrm{GeV}$, and $\stheta=0.1$.}
\end{figure}

Now one may argue that it may be better to use a combination with two different baselines
in the fashion of \Ref~\cite{Burguet-Castell:2002qx} (NF plus SB). 
Therefore, we show in \figu{twobase} the
two-baseline optimization of the NF and SB with uncorrelated matter effect
and a fixed muon energy of $E_\mu = 5 \, \mathrm{GeV}$. The optimum turns out to be close to
the same-baseline diagonal. However, it is somewhat worse than for the correlated matter density
case above. Since a option requiring only one site and, possibly, only one detector is much more
cost efficient, we
will not discuss the two-baseline case anymore. We have, in addition, tested a $500 \, \mathrm{kt}$
water Cherenkov detector for the SB part together with the $50 \, \mathrm{kt}$ {\sf Golden*}. 
More or less, we obtain the same qualitative results with the difference of
a much higher fraction of $\deltacp$: CP violation can be measured for about $91\%$ of all values of
$\deltacp$ (for $\stheta=0.1$). Of course, this option would require an additional very large detector.
Furthermore, we have checked that our optimization results to not change for somewhat smaller
$\stheta = 0.05$. In the following section, we will use these standard setups from the optimization discussion
above as given in \Tab~\ref{tab:setups}.

\section{Comparison of Facilities}

\begin{table*}[t!]
\begin{center}
\begin{tabular}{lllrc}
\hline
\\[-0.3cm]
Setup name & Description/Parameters & Det. & $m_{\mathrm{Det}}$ & \Ref \\
\hline
\\[-0.3cm]

{\sf WBB$_{\mathsf{WC}}$} & $E_p = 28 \, \mathrm{GeV}$, $P_{\mathrm{target}} = 4 \, \mathrm{MW}$ & WC & $500 \, \mathrm{kt}$ & \cite{Barger:2006vy} \\
{\sf Low-E NF} & $E_\mu=5 \, \mathrm{GeV}$, $L=900 \, \mathrm{km}$, $10^{21}$ umd & 
{\sf Golden*} & $50 \, \mathrm{kt}$ & this \\
{\sf Low-E NF\&Pt} & Same as {\sf Low-E NF}, plus platinum channel & 
 {\sf Golden*} & $50 \, \mathrm{kt}$ & this \\
{\sf High-E NF$_{\mathsf{2B}}$} & $E_\mu=20 \, \mathrm{GeV}$, $L=4\, 000 \, \mathrm{km}+7 \, 500 \, \mathrm{km}$, 
$10^{21}$ umd & 
{\sf Golden*} & 2 x $50 \, \mathrm{kt}$ & \cite{Huber:2006wb} \\
{\sf NF-SB} & $E_\mu=5 \, \mathrm{GeV}$, $L=1 \, 250 \, \mathrm{km}$, $0.5 \, 10^{21}$ umd, & {\sf Golden*} &
$50 \, \mathrm{kt}$ & this \\
& $E_p = 28 \, \mathrm{GeV}$, $P_{\mathrm{target}} = 4 \, \mathrm{MW}$ (SB) & &  &  \\
{\sf BB350} & $\gamma=350$, $L=730 \, \mathrm{km}$, $5.8\cdot10^{18}$ useful $^6$He & WC &  $500 \, \mathrm{kt}$ & \cite{Burguet-Castell:2005pa}  \\
&  decays/yr, $2.2\cdot 10^{18}$ useful $^{18}$Ne decays/yr \\ 
\hline
\end{tabular}
\end{center}
\caption{\label{tab:setups} Setups used for the comparison. The running time
is five years for each polarity for all setups. The abbreviation ``umd''
stands for ``useful muon decays per year'', the abbreviation ``WC'' for a water Cherenkov
detector, and the label {\sf Golden*} for our optimized detector (see description in \Sec~\ref{sec:setup} and \Ref~\cite{Huber:2006wb}). }
\end{table*}

\begin{figure*}[t!]
\begin{center}
\includegraphics[width=\textwidth]{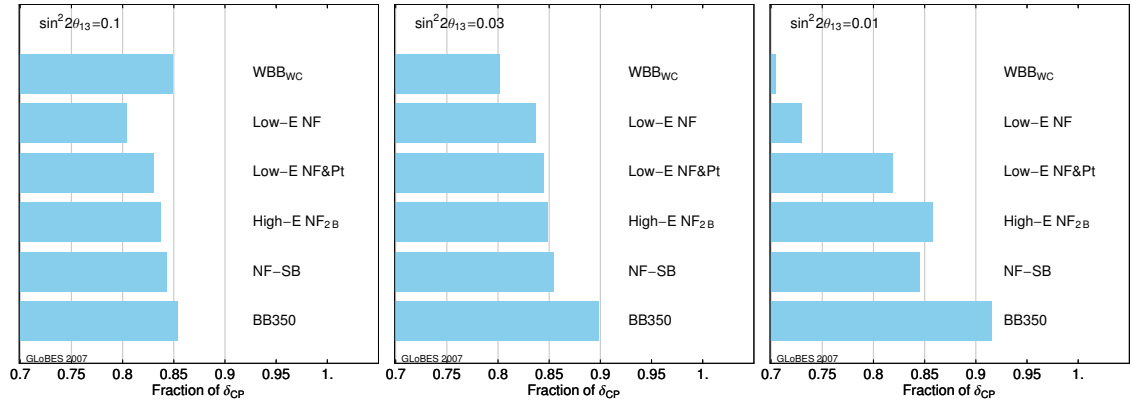}

\vspace*{-1.0cm}

\end{center}
\caption{\label{fig:comparison} Comparison of the fraction of (true) $\deltacp$
for which CP violation can be discovered ($3 \sigma$) for the different facilities
listed in \Tab~\ref{tab:setups}. The different panels correspond to different
values of $\stheta$, as given in the captions.}
\end{figure*}

We compare in \figu{comparison} the different setups from \Tab~\ref{tab:setups}
for different values of (large) $\stheta$. Obviously, the {\sf NF-SB}
outperforms any of the chosen facilities for large enough $\stheta$
except for the beta beam. In particular, it is significantly better than {\sf low-E NF} (with or without platinum channel), and slightly better than {\sf High-E NF$_{\mathsf{2B}}$}
(except for $\stheta=0.01$).  Note that these setups are operated at the 
double muon decay rate compared to the {\sf NF-SB}, which means that the information
from NF and SB is very complementary in terms of
channel and spectrum. As far as the beta beam is concerned, 
we find that it has the best absolute performance. However, given the
relatively high gamma and the challenging ion decay rates, it is difficult to judge
how this setup compares in terms of feasibility and effort. 
The comparison to the wide band beam
(in this case with a $500 \, \mathrm{kt}$ water Cherenkov detector compared to
the $50 \, \mathrm{kt}$ {\sf Golden*} used for the {\sf NF-SB} hybrid) depends very much
on $\stheta$: For very large $\stheta \simeq 0.1$, the wide band beam might be
the natural choice, because it is based on known technology. However, for a somewhat 
smaller $\stheta$, the {\sf NF-SB} is significantly better.

We have also studied further potential improvements of the {\sf NF-SB}.
Very little physics reach can be gained by the additional use of the
magic baseline or platinum channel, or by better knowledge on the
matter density profile, since the underlying correlations are
already resolved. The most significant improvement can be obtained by 
luminosity upgrades, such as by using
a large water Cherenkov detector (for the SB) in addition to {\sf Golden*} at the same site.
In this case, even the beta beam can be outperformed for very large $\stheta$.
Note that one could imagine different staged scenarios with this observation,
such as {\sf WBB$_{\mathsf{Golden*}}$} followed by {\sf NF-SB}.
We expect that further significant improvements could be obtained by
better absolute efficiencies for the muon neutrino appearance channel (for the
price of larger backgrounds), by recycling a fraction of the muons
produced in the SB channel for the NF,
or by a further optimization of proton energy and horn characterizing
the SB. 

\section{Summary and Discussion}

We have studied a low-energy neutrino factory (NF) in combination with a superbeam
(SB) produced in the decay chain of the same protons, which we have called
``neutrino factory superbeam''. We have demonstrated that this configuration outperforms
a low-energy NF alone beyond a simple addition of statistics 
even if one can use the platinum channel,
as well as it outperforms a wide band beam with the same proton energy.  
The reason for this excellent performance
is the SB spectrum peaking at lower energies, and the information
from the T-conjugated $\nu_\mu \rightarrow \nu_e$ channel at the same
baseline (\ie, with correlated matter effect). 

As far as the technical requirements for such a configuration are
concerned, the same detector might be used
at the same baseline. As for any low-energy neutrino factory, sufficiently
good efficiencies at low energies are required. The extra effort mainly reduces to a more
sophisticated target station, which has to separate the pion and kaon
charges and utilize them for the NF and SB. In addition, we have assumed that the NF and
SB bunches arrive out-of-phase at the (possibly same) detector.
For the electron neutrino detection, we have required a relatively high
efficiency without CID, which is different from
the platinum channel (where the requirement of CID significantly reduces
the efficiency). As far as systematics is concerned,
we expect this experiment to be of low risk: The NF systematics
has been assumed very conservatively, and the wide band beam is rather
robust with respect to systematics~\cite{Barger:2007jq}. This might be
an important decision criterion in comparison to narrow band beams.

From the optimization,
we have found $L \sim 800 \, \mathrm{km} - 1 \, 500 \, \mathrm{km}$ and 
$E_\mu \gtrsim 4 \, \mathrm{GeV}$ to be close-to-optional, which, for instance,
corresponds to Fermilab to Homestake or Henderson. For the proton energy,
we have tested two configurations at the lower end upper end of the admissible
range for a NF, and we have chosen $E_p = 28 \,\mathrm{GeV}$.
This means that we obtain a completely different configuration from a
high-energy NF optimized for small $\stheta$~\cite{Huber:2006wb}
in terms of baselines, muon energy, and target station. For the detector,
the need for high efficiencies at low energies is independent of
the NF configuration (but not a prerequisite for a high-energy NF). 
However, higher backgrounds may be
admissible for the low-energy NF, which means that the
overall efficiency could be higher. Although
a high-energy NF with two baselines (and possibly platinum channel)
might be used for large $\stheta$ as well, the effort might be considerably
higher. The same, in principle, applies to a higher gamma beta beam. 
We conclude that a reasonable decision can be made when the
branching point sensitivity $\stheta \simeq 0.01$ is reached. Before that
point, one may want to develop two different NF 
approaches: One optimized for $\stheta$ reach operated as a discovery machine,
the other optimized for large $\stheta$ operated as a precision instrument.

{\bf Acknowledgments}
WW acknowledges support from the Emmy Noether program of
Deutsche Forschungsgemeinschaft.
Computing was performed on facilities supported by the US NSF 
Grants EIA-032078 (GLOW), PHY-0516857 (CMS Research
Program subcontract from UCLA), PHY-0533280 (DISUN), and the
University of Wisconsin Graduate School/Wisconsin Alumni Research
Foundation.


\end{document}